\documentclass[preprint,prc,nofootinbib,A4,superscriptaddress]{revtex4-1}
\usepackage{amssymb}
\usepackage{hyperref}
\usepackage{graphicx,subfig}
\usepackage{color}
\begin{document}

\title{Hadron-quark phase transition in the neutron star with vector MIT bag model and Korea-IBS-Daegu-SKKU functional}

\author{Debashree Sen}
\affiliation{Center of Extreme Nuclear Matters, Korea University, Seoul 02841, Korea}
\author{Hana Gil}
\affiliation{Center of Extreme Nuclear Matters, Korea University, Seoul 02841, Korea}
\author{Chang Ho Hyun}
\email{hch@daegu.ac.kr}
\affiliation{Department of Physics Education, Daegu University, Gyeonsan 38453, Korea}

\date{\today}

\begin{abstract}

 Employing the Korea-IBS-Daegu-SKKU (KIDS) density functional for the hadron phase and the MIT bag model with vector (vBag) model for the quark phase, we obtain hadron-quark phase transition in neutron stars considering Maxwell construction. The structural properties of the resultant hybrid stars are computed for three different values of bag constant ($B$) in the range $B^{1/4}=$(145$-$160 MeV). We study the effects of symmetry energy ($J$) on the hybrid star properties with the different KIDS model and found that $J$ has important influence not only on the transition properties like the transition mass, transition radius and jump in density due to phase transition, but also on the stability of the hybrid stars. The vector repulsion of the quark phase via the parameter $G_V$ has profound influence in obtaining reasonable hybrid star configurations, consistent with the recent astrophysical constraints on the structural properties of compact stars. Within the aforesaid range of $B$, the value of $G_V$ is constrained to be 0.3 $\lesssim G_V \lesssim$ 0.4 in order to obtain reasonable hybrid star configurations. 

\end{abstract}

\maketitle

\section{Introduction}
\label{Intro}

Mass-radius ($M-R$) diagram of the neutron star is unprecedentedly crowded with modern astronomical data that covers the neutron star mass from about $0.5 M_\odot$ to more than $2.0 M_\odot$. They consist of GW170817 by the LIGO/Vigro Collaboration \cite{LIGOScientific:2018cki}, PSR J0030+0451 by NICER \cite{Riley:2019yda,Miller:2019cac} and PSR J0740+6620 \cite{Fonseca:2021wxt,Miller:2021qha,Riley:2021pdl} and HESS J1731-347 \cite{Doroshenko:2022}. Simultaneous measurement of the mass and radius for a single object helps greatly deepen our understanding of the state of matter at densities above the nuclear saturation.

A naive but quite certain scenario is the phase transition from hadronic matter to quark matter. Initially the matter is composed of nucleons at low densities, where they are far apart. Since the nucleon has a finite size, they start to overlap to each other at a density higher than the saturation. As the density evolves furthermore, confinement of quarks in the initial nucleon becomes uncertain, and finally quarks are completely released from the nucleons and form a deconfined quark matter. The key points to the hadron-quark phase transition are the physical properties of the nucleon and quarks, and their interactions at finite densities. The main concern of the present work is to examine the effect of the uncertainties in the interactions of quarks and nucleons to the equation of state (EoS) and the phase transition in the core of neutron stars, forming hybrid stars. For the calculation of the EoS, we use the MIT bag model with vector repulsion (vBag) \cite{Lopes:2021jpm, Kumar:2022byc, Laskos-Patkos:2023tlr, Kumar:2023lhv} for the quark matter, and the Korea-IBS-Daegu-SKKU (KIDS) density functional \cite{Gil:2020wqs, Gil:2020wct, Gil:2021ols} for the hadron phase. Phase transition is achieved with the help of Maxwell construction assuming that the surface tension of the interface is high enough ($>$70 MeV fm$^{-2}$) to ensure that the transition occurs at a sharp interface \cite{Maruyama:2007ss}. However, the value of the surface tension at the hadron-quark interface is still not well defined. Therefore, when it is $<$70 MeV fm$^{-2}$ the formation of mixed phase is also possible following Gibbs construction. Other mechanisms like hadron-quark crossover \cite{Qin:2023zrf,Huang:2022mqp,Constantinou:2021hba,Sotani:2023zkk} and constant speed of sound parameterization \cite{Sun:2023glq,Laskos-Patkos:2023cts} etc. are also adopted in literature to obtain the properties of hybrid stars.

In the vBag model, stiffness of the quark matter is controlled by the bag constant and the vector repulsion \cite{Lopes:2021jpm,Kumar:2022byc,Laskos-Patkos:2023tlr,Kumar:2023lhv}. The two quantities have different physical origins and implications. Bag constant is introduced in the original bag model. In the simplest description of the baryon bags, quarks are treated as free particles within a bag. In order to confine the quarks within the baryon, inward pressure is assumed to be exerted on the quarks. The inward pressure is described in terms of the bag constant, so it is a natural way to determine the bag constant from the baryon masses in free space. In this work, as a lower limit of the bag constant, we adopt $B^{1/4}=145$ MeV as an upper limit of the bag from Ref. \cite{Johnson:1975zp}. As far as the baryon spectrum in free space is concerned, there is no upper limit to the bag constant. However, if one adopts the Bodmer-Witten conjecture which states that the quark  matter composed of $u$, $d$ and $s$ quarks is the most stable state of matter in the quark stars, an upper limit value can be determined. In case of hybrid stars, it is not mandatory to satisfy the Bodmer-Witten conjecture and in this case the only possible way to constrain $B$ is to test the structural properties of hybrid stars in the light of the various astrophysical constraints \cite{Nandi:2017rhy,Rather:2020mlz}. We use $B^{1/4}=160$ MeV as upper limit, following Ref. \cite{Rather:2020mlz}. In this work, uncertainty arising from the bag constant is accounted by considering three values, the lower limit 145 MeV, a middle value 155 MeV and the upper limit 160 MeV.

While the bag constant is inherent in the bag model, there is no repulsive force in the original bag model. The reason might be that it is sufficient to produce the baryon spectrum accurately with the kinetic energy of quarks and the bag constant. With these two ingredients, however, it is hard to satisfy the observation of large mass stars, $M \geq 2 M_\odot$ because the EoS is too soft to sustain the strong gravity of super dense matter \cite{Lopes:2021jpm}. Request for more stiff EoS demands a repulsive interaction in terms of the vector field. Nowadays, it is common to include vector repulsion in the widely used quark models such as Nambu-Jona-Lasino model as well as the bag model. An issue related to the vector force is how strong the repulsion is. Thanks to the improvement in the precision and diversity of the neutron star observational data, it becomes feasible to constrain the strength of vector repulsion more precisely. Combined with the uncertainties of the bag constant and the symmetry energy, we can propose values of the vector coupling constant that are compatible with the observed neutron star properties.

In the description of neutron-rich hadronic matter, the largest uncertainty comes from the symmetry energy. Its value at the saturation density is relatively accurately determined, most conservatively in the range 30$-$34 MeV, but its density dependence is much more uncertain. The average density of a neutron star whose mass is larger than the canonical mass is well above the saturation density, so the neutron star provides a dependable way to determine the density dependence of the symmetry energy. In recent works, symmetry energy has been constrained by using the neutron star data \cite{Gil:2020wqs, Gil:2020wct, Gil:2021ols}, and the uncertainty could be reduced substantially. Nevertheless, the density dependence still has sizable ambiguity, and it can have critical impact to the transition to deconfined quark matter. In order to investigate the influence of the symmetry energy, we adopt the KIDS-A, B, C, D models \cite{Gil:2020wct}, which are determined to satisfy the nuclear properties, neutron star data and gravitational wave measurements simultaneously, and have evidently different stiffness of the symmetry energy.

In the calculation of the EoS, we treat the bag constant as a free parameter, and determine the range of vector coupling constants to satisfy the data of GW170817, NICER and HESS simultaneously for the given models of the symmetry energy. In the result, we find that the critical density at which the phase transition occurs is sensitive the bag constant, vector coupling constant and the symmetry energy. However, maximum mass depends on the bag constant and the symmetry energy weakly, 
so we can obtain appreciably model-independent range of the vector coupling constant.

We organize the work in the following order. In Sect. II, models for the quark and hadronic matters are described. Section III shows the results and we present the discussions on them. The work is summarized in Sect. IV. 

\section{Model}

\subsection{MIT bag with vector repulsion}
\label{vbag}

We consider quark matter with $u$, $d$ and $s$ quarks and the electrons as a lepton. The Lagrangian for the bag model with vector repulsion is given by
\begin{eqnarray}
{\cal L} &=& \sum_{f=u,d,s} \left[\bar{\psi}_f \{ \gamma^\mu (i \partial_\mu - g_{qqV} V_\mu) - m_f \}\psi_f - B \right] \Theta (\bar{\psi}_f \psi_f)
\nonumber \\
&+& \frac{1}{2} m^2_V V_\mu V^\mu - \frac{1}{4} V_{\mu \nu}V^{\mu \nu}  + \overline{\psi}_e \big(i \gamma_{\mu} \partial^{\mu} - m_e\big) \psi_e,
\end{eqnarray}
where $V_\mu$ denotes the vector field, $g_{qqV}$ is the quark-vector meson coupling constant, $B$ is the bag constant, $V_{\mu \nu} = \partial_\mu V_\nu - \partial_\nu V_\mu$ and the Heaviside function $\Theta$=1 inside the bag. For the vector coupling constant, we assume $g_{uuV} = g_{ddV}$ and $g_{ssV}/g_{uuV}$ = 0.4. It is shown that the variation on the ratio $g_{ssV}/g_{uuV}$ does not affect the EoS as much as the bag constant or the vector coupling constant \cite{Pal:2023quk}. Since the vector meson is an isoscalar particle, it corresponds to the $\omega$ meson, so we use the $\omega$ meson mass 783 MeV for $m_V$. In the quark matter EoS, vector coupling constant contributes in the form $(g_{qqV}/m_V)^2$, so we use $G_V=(g_{uuV}/m_V)^2$ as a parameter to be constrained from the neutron star data. The bag constant is considered in the range $B^{1/4} =$ 145--160 MeV, so three values 145, 155 and 160 MeV are used in the calculation.

\subsection{KIDS functional}
\label{kids}

In the KIDS density functional framework, energy of a nucleon in nuclear matter is expanded in powers of $\rho^{1/3}$ 
where $\rho$ is the matter density as
\begin{equation}
{\cal E} = {\cal T} + \sum_{i=0} \rho^{1+i/3} (\alpha_i + \beta_i \delta^2).
\end{equation}
${\cal T}$ is the kinetic energy, and $\delta=(\rho_n-\rho_p)/\rho$ where $\rho_n$ and $\rho_p$ are the neutron and the proton density, respectively. Coefficients $\alpha_i$ are determined from the properties of symmetric nuclear matter, and $\beta_i$ are constrained by the neutron star data in the KIDS-A, B, C, D models. Incompressibility of the symmetric matter $K_0$ and the symmetry energy parameters are summarized in Tab. \ref{tab1}.
\begin{table}[h]
\begin{center}
\caption{Incompressibility $K_0$ of the symmetric nuclear matter and the symmetry energy parameters $J$, $L$ and $K_{\rm sym}$ 
in units of MeV for the KIDS-A, B, C, D models.}\protect\label{tab1}
\begin{tabular}{ccccc}
\hline
\hline
 & KIDS-A & KIDS-B & KIDS-C & KIDS-D \\ \hline
$K_0$ & 230 & 240 & 250 & 260 \\
$J$ & 33 & 32 & 31 & 30 \\
$L$ & 66 & 58 & 58 & 47 \\
$K_{\rm sym}$ & $-139.5$ & $-162.1$ & $-91.5$ & $-134.5$ \\ \hline\hline
\end{tabular}
\end{center}
\end{table}

\subsection{Phase transition}
\label{PT}

In the present work, phase transition is achieved via Maxwell construction assuming that the surface tension at the hadron-quark interface is sufficiently large \cite{Maruyama:2007ss}. Maxwell construction is based on the local charge neutrality condition which implies that both the hadronic (H) and quark (Q) phases must be individually charge neutral,

\begin{eqnarray}
q_H = 0 ; q_Q = 0.
\end{eqnarray}

Following Maxwell criteria, phase transition occurs when the pressure ($P$) and baryon chemical potential ($\mu_B$) of each phase become equal,

\begin{eqnarray}
\mu_B^H = \mu_B^Q ; P_H = P_Q.
\end{eqnarray}

In case of Maxwell construction, $\mu_B$ is continuous while there is jump in electron chemical potential $\mu_e$ at the interface between the two phases. Therefore, the transition is thus characterized by jump in density from hadronic to quark phase, pressure being constant within the interval \cite{Contrera:2016phj}. The transition point ($\mu_t,P_t$) of the hadronic and quark phases in the ($\mu_B - P$) plane corresponds to two specific transition densities - one in the hadronic phase ($\rho_t^H$) and another in the quark phase ($\rho_t^Q$). $\rho_t^H$ marks the end of pure hadronic phase while $\rho_t^Q$ denotes the starting of pure quark phase in terms of density. The jump in density is given by the difference between $\rho_t^Q$ and $\rho_t^H$.

 The resultant hybrid EoS is employed to obtain the mass $M$ and the radius $R$ of hybrid stars using the Tolman-Oppenheimer-Volkoff (TOV) equations \cite{Tolman:1939jz,Oppenheimer:1939ne} while dimensionless tidal deformability $\Lambda$ is then calculated following \cite{Hinderer:2007mb,Hinderer:2009ca}. The jump in density at the hadron-quark interface is taken care of by implementing the correction in calculation of $\Lambda$ as suggested by Ref. \cite{k2_corr}. For the outer crust region, we adopt the Baym-Pethick-Sutherland EoS \cite{Baym:1971pw} and for the inner crust, we have considered the EoS including the pasta phases \cite{Grill:2014aea}. The crust-core transition density in the present work is around 0.0055$\rho_0$.

\section{Result}
\label{Results}

\begin{figure}[!ht]
\centering
{\includegraphics[width=0.5\textwidth]{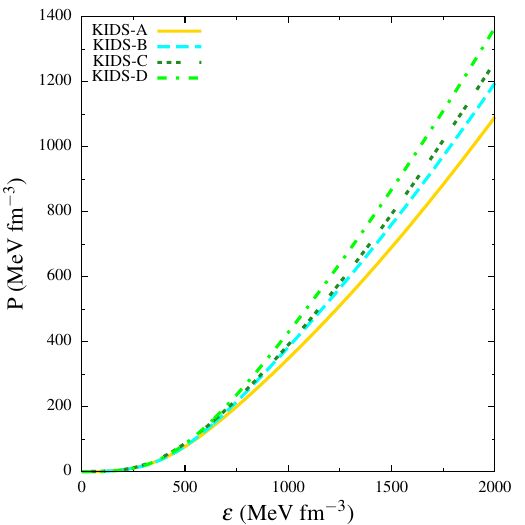}}
\caption{\it Equation of state of hadronic matter in neutron star with the four KIDS functionals.}
\label{eos}
\end{figure}

\begin{figure}[!ht]
\centering
\subfloat[]{\includegraphics[width=0.49\textwidth]{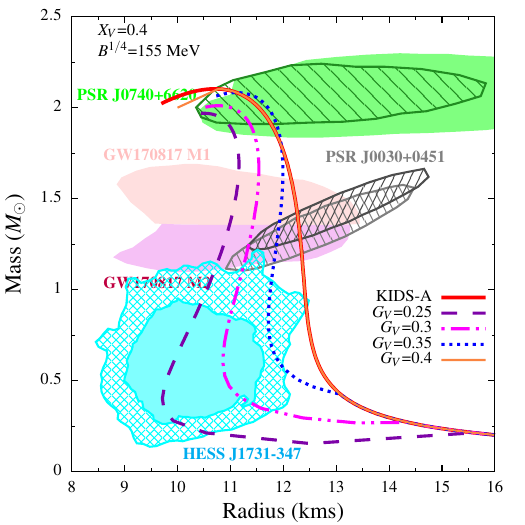}\protect\label{mrA}}
\hfill
\subfloat[]{\includegraphics[width=0.49\textwidth]{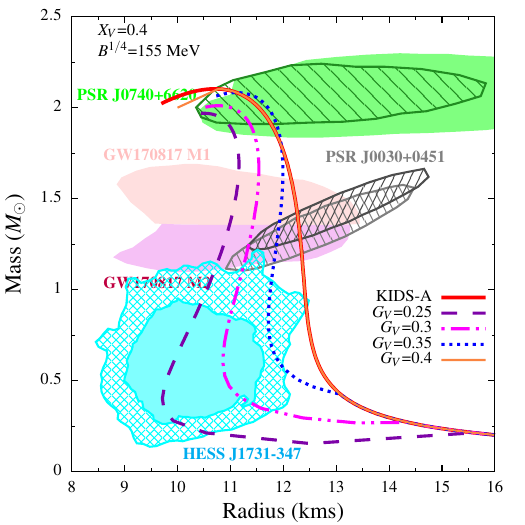}\protect\label{mrB}} \\
\subfloat[]{\includegraphics[width=0.49\textwidth]{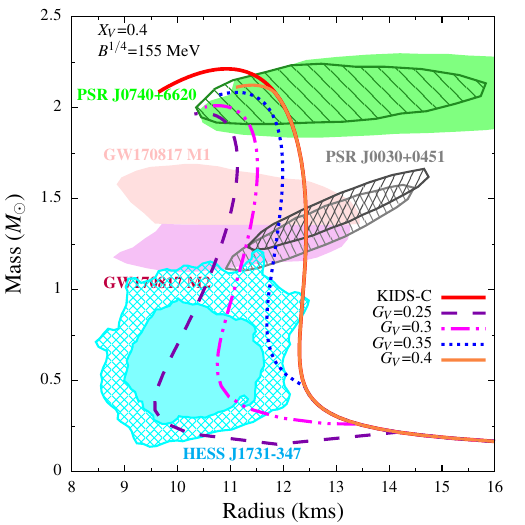}\protect\label{mrC}}
\hfill
\subfloat[]{\includegraphics[width=0.49\textwidth]{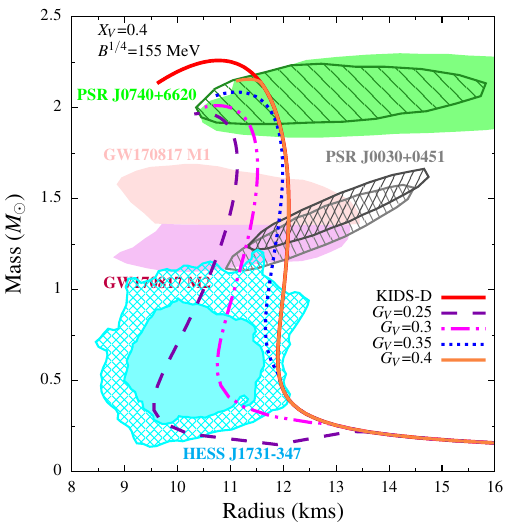}\protect\label{mrD}}
\caption{\it Variation of mass with radius of hadronic star (solid lines) and hybrid stars (dashed lines) with hadronic models (a) KIDS-A, (b) KIDS-B, (c) KIDS-C and (d) KIDS-D for different values of $G_V$ and $B^{1/4}=$155 MeV. Observational limits imposed from the most massive pulsar PSR J0740+6620 on maximum mass \cite{Fonseca:2021wxt} and corresponding radius \cite{Miller:2021qha,Riley:2021pdl} are also indicated. The constraints on $M-R$ plane prescribed from GW170817 \cite{LIGOScientific:2018cki}, NICER experiment for PSR J0030+0451 \cite{Riley:2019yda,Miller:2019cac} and HESS J1731-347 \cite{Doroshenko:2022} are also compared.}
\label{mr}
\end{figure}

We first examine the dependence of symmetry energy on the properties of hadronic stars with the four different KIDS models with different symmetry energy as tabulated in Tab. \ref{tab1}. For the hadronic model, as the symmetry energy ($J$) and its slope ($L$) decrease and compression modulus $K_0$ increases from KIDS-A to KIDS-D in Tab. \ref{tab1}, the EoS stiffens (as seen from Fig. \ref{eos}) and hence the maximum mass increases. For example, the maximum mass for KIDS-A model ($J$=33 MeV, $L$=66 MeV and $K_0$=230 MeV) is $2.10 M_{\odot}$ while it is $2.26 M_\odot$ for the KIDS-D model ($J$=30 MeV, $L$=47 MeV and $K_0$=260 MeV). Consistent with the results of \cite{Li:2020ias} we find that the radius of canonical neutron star ($R_{1.4}$) also increases with the increase of $L$ e.g., $R_{1.4}$=12.28 km for KIDS-A model and $R_{1.4}$=12.10 km for KIDS-D model. From Fig. \ref{mr} it can be seen that the hadronic star configurations with all the four KIDS model satisfy the present day astrophysical constraints on the mass-radius relation of compact stars viz., the maximum mass constraint from PSR J0740+6620 \cite{Fonseca:2021wxt} with corresponding radius constraint \cite{Miller:2021qha,Riley:2021pdl}, the constraints from GW170817 \cite{LIGOScientific:2018cki}, NICER experiment for PSR J0030+0451 \cite{Riley:2019yda,Miller:2019cac} and HESS J1731-347 \cite{Doroshenko:2022}.

\begin{figure}[!ht]
\centering
{\includegraphics[width=0.5\textwidth]{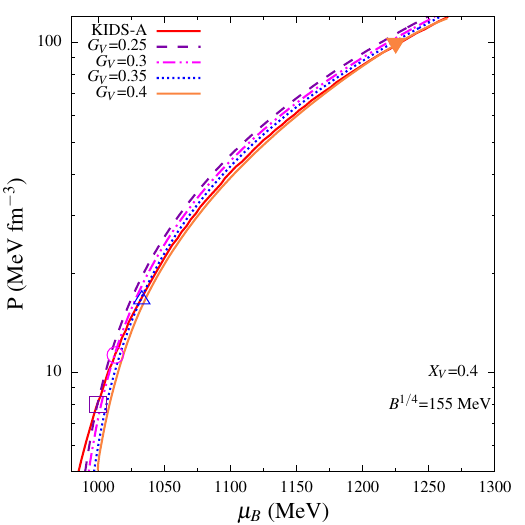}}
\caption{\it Variation of pressure as a function of baryon chemical potential for hadronic matter with KIDS-A model and quark matter for different values of $G_V$ with $X_V=$0.4 and $B^{1/4}=$155 MeV. The transition points are also marked with points.}
\label{muP}
\end{figure}

In order to study the influence of symmetry energy on the phase transition and the hybrid star structure, we consider the vBag model with $B^{1/4}$=155 MeV. In Fig. \ref{muP} we illustrate the variation of pressure as a function of baryon chemical potential of the hadronic and quark phase for example with the KIDS-A model. The transition points ($\mu_t,P_t$) are also marked. The corresponding $\rho_t^H$ and $\rho_t^Q$ are tabulated in Tab. II. We find that phase transition is quite early for low values of $G_V$ but the transition point shifts abruptly to high chemical potential and pressure (density) at $G_V$=0.4. This is consistent with the result obtained by Lopes {\it et al.} \cite{Lopes:2021jpm}.
As a result we observe in Fig. \ref{mr} that for any particular hadronic model, the maximum mass and corresponding radius of the hybrid stars increase with increasing repulsion via $G_V$. The value of $R_{1.4}$ and the radius of low mass hybrid stars are greatly affected by $G_V$. Fig. \ref{mr} also confirms that all the present day astrophysical constraints are satisfied by the hybrid star configurations with all the four KIDS model with the chosen values of $G_V$ except for 0.25, for which the NICER data for PSR J0030+0451 are not satisfied. Also for any particular hadronic model, the recently obtained HESS J1731-347 data for different values of $G_V$ with $X_V$=0.4 and $B^{1/4}$=155 MeV is better satisfied with lower repulsion i.e., lower values of $G_V$. The transition mass ($M_t$) increases while the transition radius ($R_t$) decreases with increasing values of $G_V$, and for $G_V$=0.4 there is an abrupt jump in the value of $M_t$ for all the hadronic models as a consequence of the delayed transition. Such delayed transition leads to unstable hybrid star configurations obtained with both KIDS-A and KIDS-B hadronic models as seen from Figs. \ref{mrA} and \ref{mrB}.

\begin{figure}[!ht]
\centering
\subfloat[]{\includegraphics[width=0.49\textwidth]{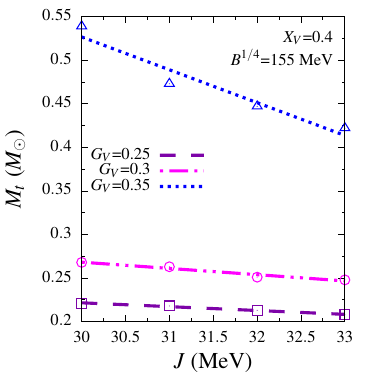}\protect\label{MtJ}}
\subfloat[]{\includegraphics[width=0.49\textwidth]{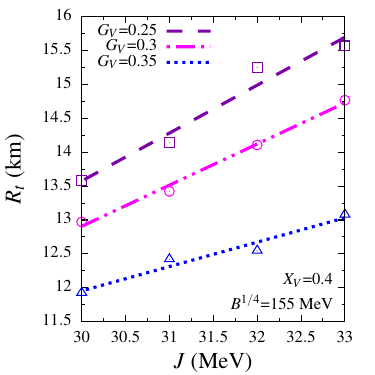}\protect\label{RtJ}}
\caption{\it Variation of (a) transition mass and (b) transition radius of hybrid stars with respect to symmetry energy for different values of $G_V$  with $X_V=$0.4 and $B^{1/4}=$155 MeV.}
\label{MtRt_J}
\end{figure}

\begin{figure}[!ht]
\centering
{\includegraphics[width=0.5\textwidth]{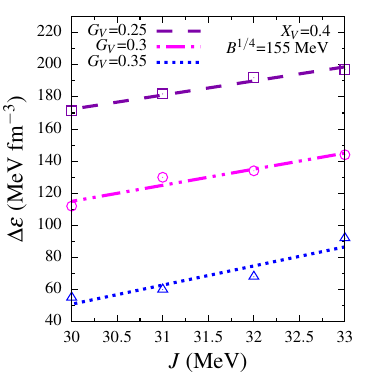}}
\caption{\it Variation of jump in energy density of hybrid stars with respect to symmetry energy for different values of $G_V$ with $X_V=$0.4 and $B^{1/4}=$155 MeV.}
\protect\label{delE_J}
\end{figure}

 We show the dependence of transition properties like $M_t$, $R_t$ and the energy difference ($\Delta \varepsilon=\varepsilon_t^Q -\varepsilon_t^H$) 
or the jump in density from hadronic to quark phase as functions of symmetry energy $J$ in Figs. \ref{MtRt_J} and \ref{delE_J}. We do not show the case for $G_V$=0.4 since it leads to unstable hybrid star configurations for high values of $J$. For any particular value of $G_V$, the relationships (fitted) between $M_t$ and $R_t$ with $J$ show that symmetry energy has a considerable influence on the transition properties of hybrid stars. From Fig. \ref{MtRt_J} we find that for increasing (decreasing) values of $J$ ($K_0$) the transition mass decreases considerably (Fig. \ref{MtJ}) while the transition radius follows the reverse trend (Fig. \ref{RtJ}). The relation between $\Delta \varepsilon$ and $J$ in Fig. \ref{delE_J} shows that $\Delta \varepsilon$ also increases with increasing $J$ or decreasing $K_0$ for any particular value of $G_V$. The dependence of $\Delta \varepsilon$ on $J$ is almost linear and the slope is almost independent of the value of $G_V$.

\begin{figure}[!ht]
\centering
\subfloat[]{\includegraphics[width=0.49\textwidth]{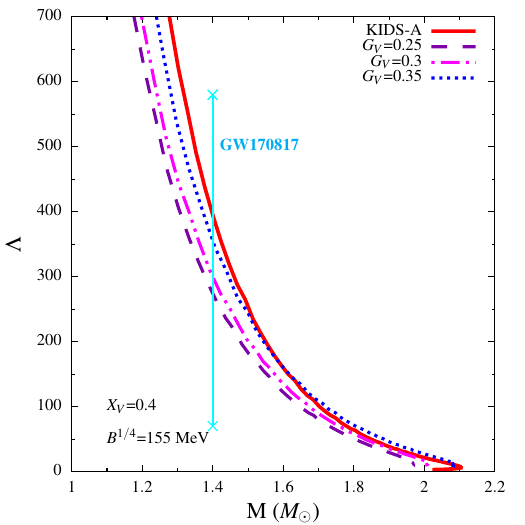}\protect\label{LamA}}
\hfill
\subfloat[]{\includegraphics[width=0.49\textwidth]{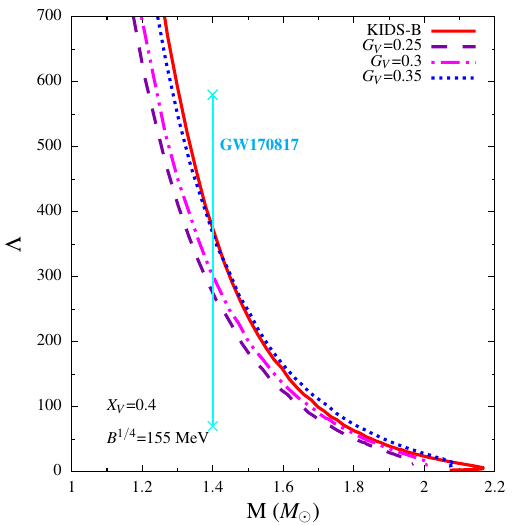}\protect\label{LamB}} \\
\subfloat[]{\includegraphics[width=0.49\textwidth]{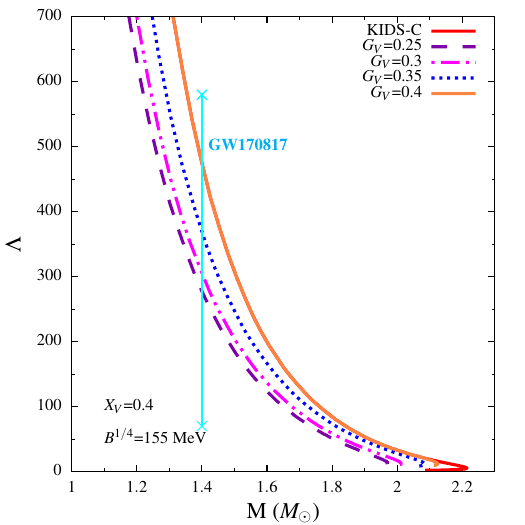}\protect\label{LamC}}
\hfill
\subfloat[]{\includegraphics[width=0.49\textwidth]{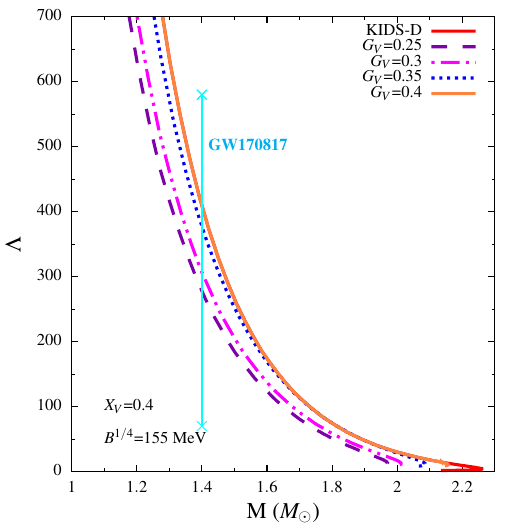}\protect\label{LamD}}
\caption{\it Variation of tidal deformability with mass of hadronic star (solid lines) and hybrid stars (dashed lines) with hadronic models 
(a) KIDS-A, (b) KIDS-B, (c) KIDS-C and (d) KIDS-D for different values of $G_V$ and $B^{1/4}=$155 MeV.}
\label{Lam}
\end{figure}

We also study the variation of the dimensionless tidal deformability with the canonical mass of hadronic and hybrid stars in Fig. \ref{Lam}, 
where it is seen that the values of $\Lambda_{1.4}$ for all the hadronic and hybrid star configurations are in good agreement with the constraint from GW170817 \cite{LIGOScientific:2018cki}. However, dependence on the $G_V$ value is evident, so 
$\Lambda_{1.4} \simeq 260$ with $G_V$=0.25 but $\Lambda_{1.4}\simeq$ 380 with $G_V$=0.35. Accurate measurement of the tidal deformability will certainly help determine $G_V$ value less ambiguously.

\begin{figure}[!ht]
\centering
\subfloat[]{\includegraphics[width=0.49\textwidth]{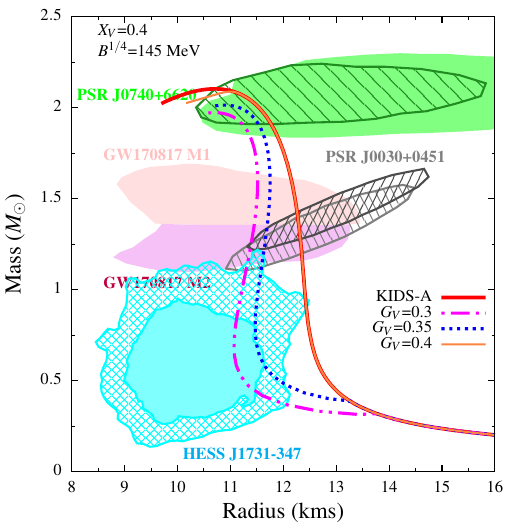}\protect\label{mrA145}}
\subfloat[]{\includegraphics[width=0.49\textwidth]{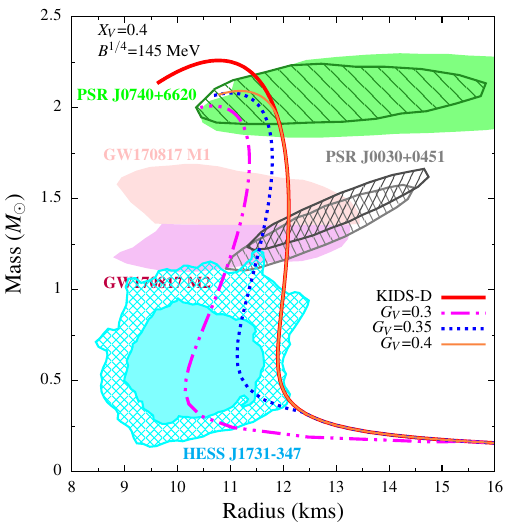}\protect\label{mrD145}} \\
\subfloat[]{\includegraphics[width=0.49\textwidth]{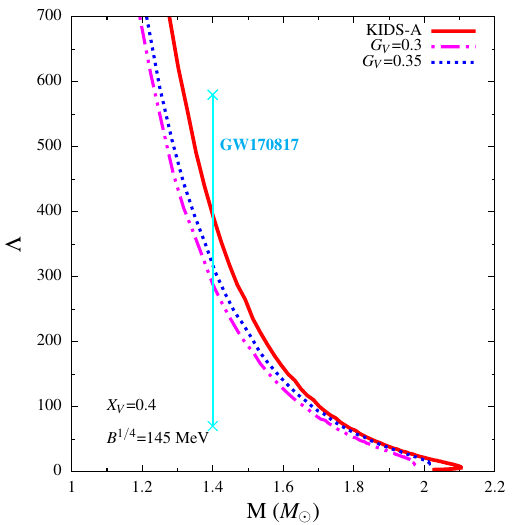}\protect\label{LamM145_A}}
\subfloat[]{\includegraphics[width=0.49\textwidth]{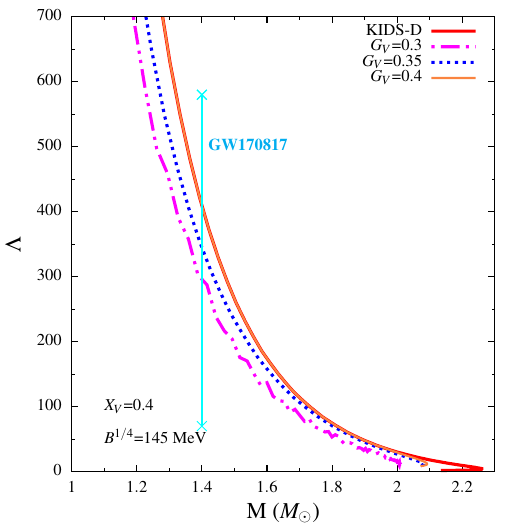}\protect\label{LamM145_D}}
\caption{\it Variation of mass with radius (a, b) and tidal deformability with mass (c, d) of hadronic star (solid lines) and hybrid stars (dashed lines) 
with hadronic models KIDS-A and KIDS-D for different values of $G_V$ and $B^{1/4}=$145 MeV.}
\label{mrLam145}
\end{figure}

We next consider a lower value of $B$ as $B^{1/4}=$145 MeV. We have already seen from Fig. \ref{mr} that for $B^{1/4}=$155 MeV, none of the hybrid star configurations obtained with $G_V$=0.25 and the four KIDS model could satisfy the NICER data for PSR J0030+0451 due to early transition. Therefore, it can be expected that further low value of $B$ will lead to further early or no transition at all. For $B^{1/4}=$145 MeV we do not consider $G_V$=0.25. We, however, consider $G_V$=0.4 to check if early transition for lower value of $B$ can give rise to stable hybrid stars. For the purpose we consider the softest (KIDS-A) and the stiffest (KIDS-D) among the four KIDS models. As expected, phase transition is early in the case of $B^{1/4}$=145 MeV. This is also reflected in Fig. \ref{mrA145} and \ref{mrD145} where it can be seen that the transition mass is lower than that in case of $B^{1/4}=$155 MeV for a particular value of $G_V$. For example, considering the KIDS-A model ($J$=33 MeV) and $G_V$=0.35, $M_t$=0.422 $M_{\odot}$ when $B^{1/4}$=155 MeV as seen from Figs. \ref{MtJ} and \ref{mrA} while $M_t$=0.389 $M_{\odot}$ when $B^{1/4}=$145 MeV as seen from Fig. \ref{mrA145}. Considering KIDS-D model ($J$=30 MeV) and $G_V$=0.35, $M_t$=0.539 $M_{\odot}$ as seen from Figs. \ref{MtJ} and \ref{mrD} for $B^{1/4}$=155 MeV while $M_t$=0.397 $M_{\odot}$ in case of $B^{1/4}=$145 MeV as seen from Fig. \ref{mrD145}. Simultaneously, the transition radius is also affected by $B$. For example, in case of KIDS-A and $G_V$=0.35, $R_t$=13.081 km for $B^{1/4}$=155 MeV as seen from Figs. \ref{RtJ} and \ref{mrA} while $R_t$=13.255 km in case of $B^{1/4}$=145 MeV as seen from Fig. \ref{mrA145}. Considering the KIDS-D model and $G_V$=0.35, $R_t$=11.924 km as seen from Figs. \ref{RtJ} and \ref{mrD} for $B^{1/4}$=155 MeV while $R_t$=12.171 km in case of $B^{1/4}$=145 MeV as seen from Fig. \ref{mrD145}. Therefore transition mass is lowered while transition radius is increased for lower value of $B$. Similar to the case of higher value of $B^{1/4}$=155 MeV, the hybrid star configurations obtained for $G_V$=0.4 with $B^{1/4}$=145 MeV is unstable for the KIDS-A model as seen from Figs. \ref{mrA} and \ref{mrA145}. It is, however, stable with both the values of $B$ and $G_V$=0.4 in case of KIDS-D model having value of slope of symmetry energy $L$ much lower than that of the KIDS-A model as reflected in Tab. \ref{tab1}. So, it can be said that slope of symmetry energy $L$ also affects the stability of hybrid stars and EoS with lower value of the symmetry energy is more suitable for obtaining stable hybrid star configurations with large quark repulsion $G_V>$ 0.35. We also find that for $B^{1/4}$=145 MeV, all the present day astrophysical constraints are satisfied with $G_V$=0.3, 0.35 and 0.4 in terms of mass, radius (Figs. \ref{mrA145} and \ref{mrD145}) and tidal deformability (Figs. \ref{LamM145_A} and \ref{LamM145_D}). Another notable result is, similar to the result of $B^{1/4}=155$ MeV, there is a huge gap in the transition mass and transition density between $G_V=0.35$ and $G_V$=0.4. The transition mass is close to the maximum mass for $G_V$=0.4, which means that this delayed transition do not let the maximum mass of the hybrid star increase substantially above the transition mass.

\begin{figure}[!ht]
\centering
\subfloat[]{\includegraphics[width=0.49\textwidth]{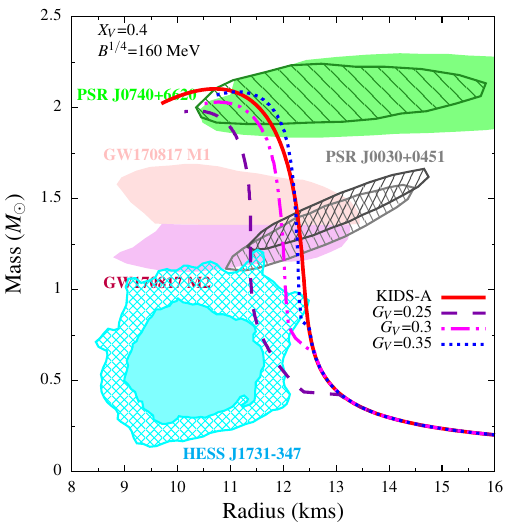}\protect\label{mrA160}}
\subfloat[]{\includegraphics[width=0.49\textwidth]{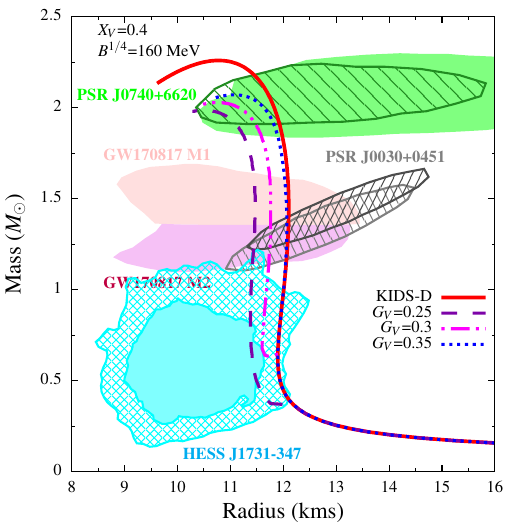}\protect\label{mrD160}} \\
\subfloat[]{\includegraphics[width=0.49\textwidth]{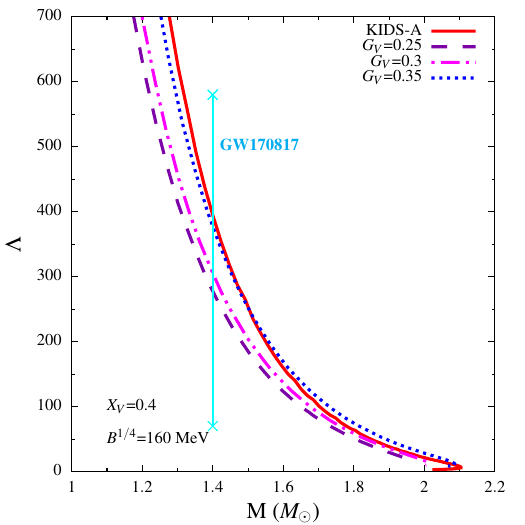}\protect\label{LamM160_A}}
\subfloat[]{\includegraphics[width=0.49\textwidth]{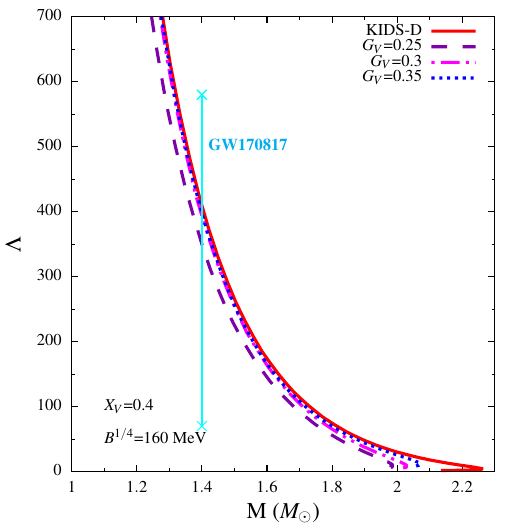}\protect\label{LamM160_D}}
\caption{\it Variation of mass with radius (a, b) and tidal deformability with mass (c, d) of hadronic star (solid lines) and hybrid stars (dashed lines) with hadronic models KIDS-A and KIDS-D for different values of $G_V$ and $B^{1/4}=$160 MeV.}
\label{mrLam160}
\end{figure}

We finally consider a higher value of $B$ as $B^{1/4}$=160 MeV which we consider as the upper limit of $B$ to obtain hybrid star configurations 
with the KIDS-A and KIDS-D hadronic models. 
With this choice of $B$, it is expected that for $G_V$=0.4, transition will be further delayed and since we already obtained unstable hybrid star configurations with lower values of $B$, we do not consider $G_V$=0.4 in case of the maximum value of $B$. However, with low values of $G_V$ in case of $B^{1/4}$=160 MeV, transition density may be suitable to achieve reasonable hybrid star configurations. Therefore we also consider $G_V$=0.25 along with 0.3 and 0.35 in this case. The results are displayed in Fig. \ref{mrLam160}. We find from Figs. \ref{mrA160} and \ref{mrD160} that even low quark repulsion ($G_V$=0.25) can satisfactorily satisfy all the astrophysical and observational constraints including the NICER data for PSR J0030+0451 which was not satisfied with the lower values of bag constant. Considering $G_V$=0.35 and $B^{1/4}$=160 MeV, $M_t$=0.798$M_\odot$ and 0.963$M_{\odot}$ while $R_t$=12.479 km and 11.905 km for the KIDS-A and KIDS-D models, respectively. It can thus be concluded that for a particular value of $G_V$ and $J$, $M_t$ increases with $B$ while $R_t$ follows the reverse trend. Also, for a particular value of $G_V$ and $B$, $M_t$($R_t$) decreases(increases) with a larger $J$. However, the variation of $J$ is quite small compared to that of $L$ in the case of the four KIDS model considered in the present work. So it can be said that the transition and hybrid star properties are more sensitive to the value of $L$. Summarizing the results up to now, depending on the value of the bag constant and the symmetry energy, the hybrid star configurations for $G_V$ values in the range $0.25 \lesssim G_V \lesssim 0.4$ are consistent with the observational data.

In Tab. \ref{table_trans} we display the transition densities and the maximum mass of the various hybrid star configurations. At the saturation density $\rho_0$=0.16 fm$^{-3}$, the volume occupied by a nucleon is $V$=1/$\rho_0$=6.25 fm$^3$. Assuming that the nucleon is a sphere located at the center of a cube whose volume is 6.25 fm$^3$, the distance between the center of two nucleons is 1.84 fm. The radius of a nucleon is about 0.8 fm, so there is no overlapping between the nucleons at the saturation density. When the distance between the center of two nucleons is 1.6 fm, the surface of a nucleon touches the surface of other nucleons. The volume of a cube with the length of one side 1.6 fm is 4.10 fm$^3$, and it gives a density $\rho$=0.244 fm$^{-3}$ = 1.53$\rho_0$. Taking into account the swelling of nucleons in nuclear medium, onset of the overlapping could happen at densities smaller than $1.5 \rho_0$, but it must be larger than the saturation density. Considering this mechanical condition for the phase transition, $G_V$=0.25 is ruled out in any combinations of $B$ and the symmetry energy. Even $G_V$=0.3 is not acceptable for some cases. Therefore, combining the consistency with the astronomical data and the mechanical condition for the phase transition, acceptable range of $G_V$ is $0.3 \lesssim G_V \lesssim 0.4$. Since for high values of $J$, $G_V$=0.4 yields unstable hybrid star configurations and for high values of $B$ there is no transition at all, $G_V$=0.4 could be regarded as an upper limit of $G_V$ value if one assumes the phase transition in the neutron star.

\clearpage

\begin{table*}[!ht]
\begin{center}
\caption{Comparison of hadron-quark transition densities, and maximum mass for different $G_V$ with different values of $B$.}
\setlength{\tabcolsep}{10.0pt}
{\small{
\begin{center}
\begin{tabular}{cccccccc}
\hline
\hline
\multicolumn{1}{c}{$B^{1/4}$} &
\multicolumn{1}{c}{Hadronic Model} &
\multicolumn{1}{c}{$G_V$} &
\multicolumn{1}{c}{$\rho_t^H/\rho_0$} &
\multicolumn{1}{c}{$\rho_t^Q/\rho_0$} &
\multicolumn{1}{c}{$M_{max}$}  \\
\multicolumn{1}{c}{(MeV)} &
\multicolumn{1}{c}{} &
\multicolumn{1}{c}{} &
\multicolumn{1}{c}{} &
\multicolumn{1}{c}{} &
\multicolumn{1}{c}{($M_{\odot}$)}   \\
\hline
145  &KIDS-A    &0.3   &0.821  &1.875  &1.972 \\
     &          &0.35  &1.273  &1.963  &2.014 \\
     &          &0.4   &5.800  &6.238  &2.088 (unstable) \\         
\hline
     &KIDS-D    &0.3   &1.006  &1.900  &2.009  \\
     &          &0.35  &1.533  &1.983  &2.081\\
     &          &0.4   &3.845  &4.638  &2.092\\         
\hline
\hline
155  &KIDS-A     &0.25  &0.737  &1.955  &1.969  \\
     &           &0.3   &1.093  &2.218  &2.011  \\
     &           &0.35  &1.444  &2.468  &2.083  \\
     &           &0.4   &6.500  &7.262  &2.103 (unstable)\\         
\hline
     &KIDS-D     &0.25  &0.606  &1.937  &1.969  \\
     &           &0.3   &1.219  &2.401  &2.011  \\
     &           &0.35  &1.756  &2.897  &2.084   \\
     &           &0.4   &4.563  &5.137  &2.156   \\         
\hline
\hline
160  &KIDS-A     &0.25  &0.794  &1.973  &1.983  \\
     &           &0.3   &1.146  &2.344  &2.031  \\
     &           &0.35  &1.500  &2.508  &2.087  \\
\hline
     &KIDS-D     &0.25  &0.671  &1.997  &1.985 \\
     &           &0.3   &1.269  &2.088  &2.028 \\
     &           &0.35  &1.803  &2.179  &2.087 \\
\hline
\hline
\end{tabular}
\end{center}
}}
\protect\label{table_trans}
\end{center}
\end{table*} 

\newpage

\section{Summary}

The present work is dedicated to study hadron-quark phase transition and the structural properties of hybrid stars. They depend on the properties of both the hadronic and quark phases like the symmetry energy of the former and the bag constant and the strength of quark repulsion of the latter. 
In order to investigate the effects of symmetry energy on the hybrid star properties we consider four KIDS models with different symmetry energy while the vBag model is adopted for the quark phase to study the influence of quark repulsion via the different values of vector coupling $G_V$ at three individual values of bag constant $B$.

 We found that for any particular value of $G_V$ and $J$, higher bag constant leads to delayed transition in terms of both density and mass. Therefore, the maximum mass $M_{max}$ of hybrid stars also increases with increasing value of $B$. Moreover, the transition mass $M_t$ increases with $B$ while $R_t$ follows the reverse trend. 
 
For any particular value of $B$ and $J$, $M_{max}$ increases with increasing value of $G_V$. The value of $R_{1.4}$ is also greatly affected by $G_V$. The radius of low mass hybrid stars are seen to increase with increasing values of $G_V$. $M_t$ increases while $R_t$ decreases with the increase of $G_V$.

For fixed values of $G_V$ and $B$, $M_{max}$ increases with decreasing value of $J$. $M_t$($R_t$) decreases(increases) with increasing $J$. The jump in energy density $\Delta \varepsilon$ from hadronic to quark phase also increases with increasing value of $J$ for any particular value of $G_V$. The dependence (fitted) of $\Delta \varepsilon$ on $J$ is linear and the slope is almost independent of the value of $G_V$. Considering a reasonable and moderately wide range of $B^{1/4}$=(145$-$160) MeV and symmetry energy $J$=(30$-$33) MeV, we find that the most suitable value of $G_V$ is approximately 0.3 $\lesssim G_V \lesssim$ 0.4 for the hybrid stars to satisfy all the present day astrophysical constraints on the structural properties of compact stars. $G_V \leq$ 0.25 is allowed only when the bag constant is as high as $B^{1/4}>$ 155 MeV in the light of the NICER data for PSR J0030+0451. The stability of hybrid stars configurations with large quark repulsion $G_V>$ 0.35 is also affected by symmetry energy. We find that lower values of symmetry energy ($J<$ 32 MeV) are more suitable for obtaining stable hybrid star configurations with large quark repulsion $G_V>$ 0.35. Accurate measurement of the tidal deformability can make a significant contribution to reducing the uncertainty of $G_V$ in a narrower range.

In recent literature, it is shown that the consideration of extensive microscopic methods for the hadronic and quark matter suggests that the crossover transition is more likely to happen than the strong first-order transition in the core of neutron stars \cite{Qin:2023zrf,Huang:2022mqp,Constantinou:2021hba,Sotani:2023zkk}. Therefore, it will be interesting to study the possibilities of such crossover transition in neutron stars in our near future projects.

\section*{Acknowledgements}

This work was supported by the National Research Foundation of Korea (Grant Nos. 2018R1A5A1025563 and 2023R1A2C1003177).

\end{document}